# Deep Learning Stock Volatility with Google Domestic Trends

Ruoxuan Xiong[1], Eric P. Nichols[2] and Yuan Shen[*3]

[1]Department of Management Science and Engineering, Stanford University

[2]Google Inc.

[3]Department of Physics, Stanford University

**Abstract**

We have applied a Long Short-Term Memory neural network to model S&P 500 volatility, incorporating Google domestic trends as indicators of the public mood and macroeconomic factors. In a held-out test set, our Long Short-Term Memory model gives a mean absolute percentage error of 24.2%, outperforming linear Ridge/Lasso and autoregressive GARCH benchmarks by at least 31%. This evaluation is based on an optimal observation and normalization scheme which maximizes the mutual information between domestic trends and daily volatility in the training set. Our preliminary investigation shows strong promise for better predicting stock behavior via deep learning and neural network models.

## Introduction

Forecasting highly volatile financial time series, *e.g.* stock returns at an intermediate frequency, is a challenging task in the presence of strong noise. However, we believe that how noisy the noise is (*i.e.* the volatility) can be modeled with decent accuracy. In a typical setup, one has input features containing the market information as well as variables external to the market. Feature selection, observation frequency, normalization method, and the model structure together determine how good the prediction can be.

Artificial neural networks are good nonlinear function approximators [1], so they are a natural approach to consider with modeling time series which are suspected to have nonlinear dependence on inputs. It is indeed not new to forecast financial time series using machine learning methods and recurrent neural networks are well-suited to this task. For instance, this early (1990) work [2] is among the first of several which use recurrent neural nets to predict stock prices, Ref. [3] instead reported a volatility forecasting model, and Ref. [4] incorporated public mood data in directional prediction of the Dow Jones Industrial Average.

However, models may not necessarily be more effective when they become more complicated. Overfitting is one of the biggest issues which have plagued highly-parameterized supervised machine learning methods in past decades [5]. Besides, an increased degree of freedom can also cause the training process to be trapped in some local minimum of the high dimensional functional space even when the model is an honest representation of the system. Fortunately, recent advances in neural networks leverage their predictive power by providing more insight into how they operate and systematically avoiding the spectre of overfitting. Specifically, there are new regularization methods (*e.g.* "dropout" [20]) and faster training techniques such as using piecewise linear activation functions as opposed to transcendental functions [6]), which allow for neural nets with many hidden layers to be trained easily — hence the term "Deep Learning". In addition, new visualization techniques have been demonstrated [7] which give users more insight into how artificial neural networks operate. These advances have together paved the way for more effective training novel architectures such as the Long Short-Term Memory (LSTM) [15]. This particular type of recurrent neural network has shown remarkable results in tasks such as artificial handwriting generation [8], language forecasting [9], and speech recognition [10].

In this work, we attempt to predict the S&P 500 volatility using an LSTM which incorporates Google domestic trends together with market data. In the data section, we will introduce our input features such as these domestic trends, and explore the predictive power of different observation and normalization schemes. In the following method section, we will specify our LSTM model together with other benchmark models so that the results of this work can be reproduced. Finally, we will compare our model performance with the benchmarks and discuss issues including applications at other frequencies, overfitting, and error statistics.

## Data Sources

In this work, we study the S&P 500 market fund based on publicly available daily data comprising high, low, open, close, and adjusted close prices. Daily returns $r_t$ are evaluated as the log difference of the adjusted close price, while daily volatility $\sigma_t$ is estimated using the high, low, open and close prices in equ. 2 [11].

$$u = \log\left(\frac{Hi_t}{Op_t}\right), \quad d = \log\left(\frac{Lo_t}{Op_t}\right), \quad c = \log\left(\frac{Cl_t}{Op_t}\right) \quad (1)$$

$$\sigma_t = 0.511\,(u-d)^2 - 0.019\,[c\,(u+d) - 2ud] - 0.383c^2 \quad (2)$$

*sy0302@stanford.edu

Starting from the year 2004, Google has been collecting the daily volume of searches related to various aspects of macroeconomics. This database is available to the public as the Google domestic trends $d_t$. A recent study has shown correlations between Google trends and the equity market [12]. In this work, we use this trend data as a representation of the public interest in various macroeconomic factors. Fig. 1 shows an example of “Bankruptcy” scaled by the relative fraction in total Google searches on 1-Jan-2004. The marked maximum at the year 2008 corresponds to the financial crisis of 2007-08.

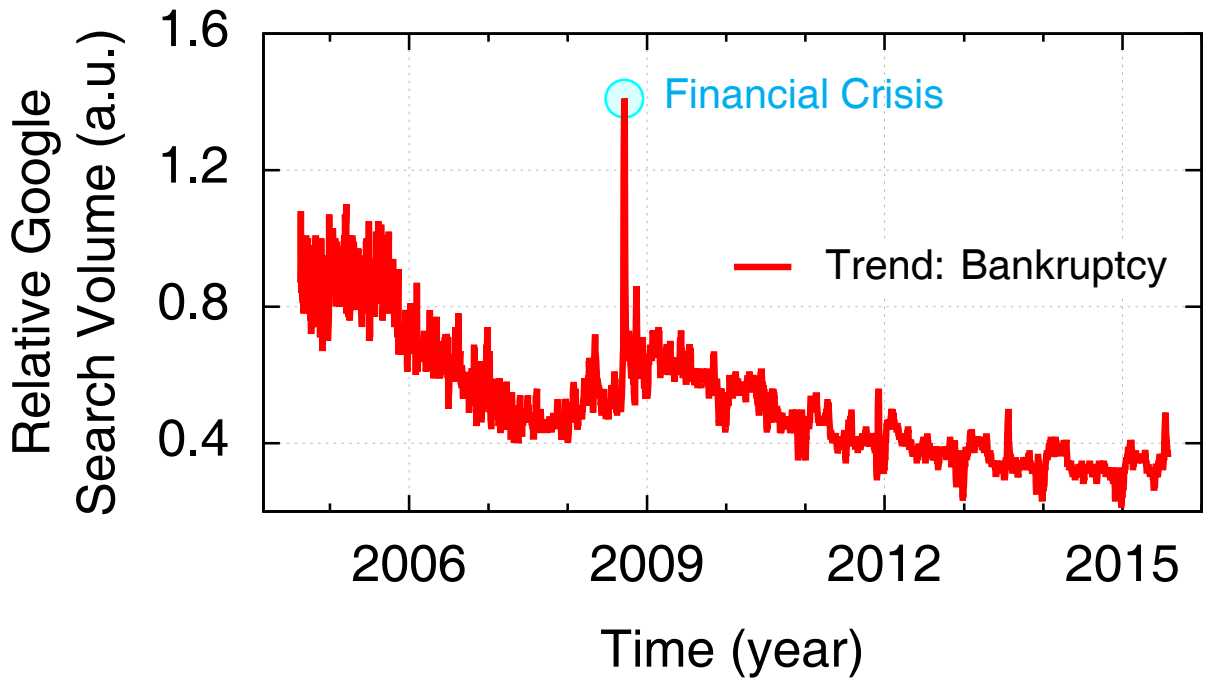

Figure 1: Relative Google Search Volume on “Bankruptcy” as a domestic trend. The data is scaled by the value at the beginning (1-Jan-2004) of the time series. The maximum appears at the time of the great financial crisis, which is highlighted by the cyan circle.

For this study, we include 25 domestic trends which are listed in Table 1 with their abbreviations.

| Trend | Abbreviation |
|---|---|
| advertising & marketing | advert |
| air travel | airtvl |
| auto buyers | autoby |
| auto financing | autofi |
| business & industrial | bizind |
| bankruptcy | bnkrpt |
| computers & electronics | comput |
| credit cards | crcard |
| durable goods | durble |
| education | educat |
| finance & investing | invest |
| financial planning | finpln |
| furniture | furntr |
| insurance | insur |
| jobs | jobs |
| luxury goods | luxury |
| mobile & wireless | mobile |
| mortgage | mrtge |
| real estate | rlest |
| rental | rental |
| shopping | shop |
| small business | smallbiz |
| travel | travel |

Table 1: Google domestic trends incorporated in this study.

Combining this data with the observed S&P 500 returns and volatility, we construct an input $x_{\lambda,t}$ of 25 dimensions. Here $\lambda$ and $t$ run over input dimension and daily time, respectively.

$$x_\lambda = \left(r, \sigma, d_{\text{advert}}, \ldots, d_{\text{travel}}\right) \tag{3}$$

We split the whole data set into a training set (70%) and a test set (30%). The training set ranges from 19-Oct-2004 to 9-Apr-2012 while the test set ranges from 12-Apr-2012 to 24-Jul-2015. Additionally, it is worth noting here that all these 25 time series are stationary in the sense that their unit-root null hypotheses have p-values less than 0.05 in the Augmented Dickey-Fuller test [13].

Preprocessing the time series with different observation and normalization schemes may result in different patterns of causality between the input and output. Both the input and output time series should be transformed from the daily data if a different scheme is chosen. Let $\Delta t$ be the observation interval,

$$r_i^{\Delta t} = \sum_{t=\left(i-1\right)\Delta t+1}^{i\Delta t} r_t \tag{4}$$

$$d_i^{\Delta t} = \frac{1}{\Delta t} \sum_{t=\left(i-1\right)\Delta t+1}^{i\Delta t} d_t \tag{5}$$

$$\sigma_i^{\Delta t} = \sqrt{\sum_{t=\left(i-1\right)\Delta t+1}^{i\Delta t} \sigma_t^2}. \tag{6}$$

In this study, volatility is studied as the output, but note that return prediction could also be done with a similar model. Since we may choose to observe the output at a different frequency other than daily, we denote the next period output $y^{\Delta t}$ as $\sigma_{+1}^{\Delta t}$ or $r_{+1}^{\Delta t}$.

$$y_i^{\Delta t} = \sigma_{i+1}^{\Delta t} \quad \text{or} \quad r_{i+1}^{\Delta t} \tag{7}$$

Normalization can be done by computing z-scores with a sliding look-back window of $k$ days for any time series $A$.

$$\mathcal{Z}_{k,i}^{A} = \frac{A_i - \text{mean}\left(A_{i-k:i}\right)}{\text{std}\left(A_{i-k:i}\right)}. \tag{8}$$

We would like to note here that $k = \infty$ corresponds to linear transformation of the time series $A$, and the mean and standard deviation within an infinitely large sliding window can be evaluated on the entire training set.

Each combination of $\Delta t$ and $k$ should determine an observation and normalization scheme with its unique predictive power. We denote these schemes as $\left(\Delta t, k\right)$. In principle, one may apply learning models on each scheme and evaluate the accuracy of prediction on a validation set such that the optimal scheme can be chosen. Alternatively, an information metric can be set up to select the optimal scheme which maximizes this metric. In this work, we use the mutual information [14] for each $\left(\Delta t, k\right)$. Assuming conditional independence between the input variables, the mutual information can be broken down into a sum of the individual components of $x_\lambda$.

$$\mathcal{MI}\left(\mathcal{Z}_k^{x^{\Delta t}}, \mathcal{Z}_k^{y^{\Delta t}}\right) = \sum_\lambda \mathcal{MI}\left(\mathcal{Z}_k^{x_\lambda^{\Delta t}}, \mathcal{Z}_k^{y^{\Delta t}}\right) \tag{9}$$

Fig. 2 shows, using the same color scheme, equ. 9 evaluated for both $y = r_{+1}$ (a) and $y = \sigma_{+1}$ (b) in the training set. The predictive power for the returns, in the sense of mutual information, is significantly smaller than that for volatility. This observation is consistent with a similar unpublished study we have done on the returns and the fact that returns are extremely noisy on the minute-to-daily timescale. Although the noise is unpredictable, how noisy the noise is may be tractable. Fig. 2 gets darker as $\Delta t$ increases and presumably the mutual information finds its maximum value when $\Delta t$ approaches $\infty$. This is a timescale in which all noises are essentially averaged out and one is left alone with a deterministic drift. Doing the normalization, on the other hand, can either reduce non-stationarity or increase the noise-to-signal ratio. The competition between these two factors results in a local maximum of the mutual information close to $k = 30$, $t = 6$ in fig. 2b.

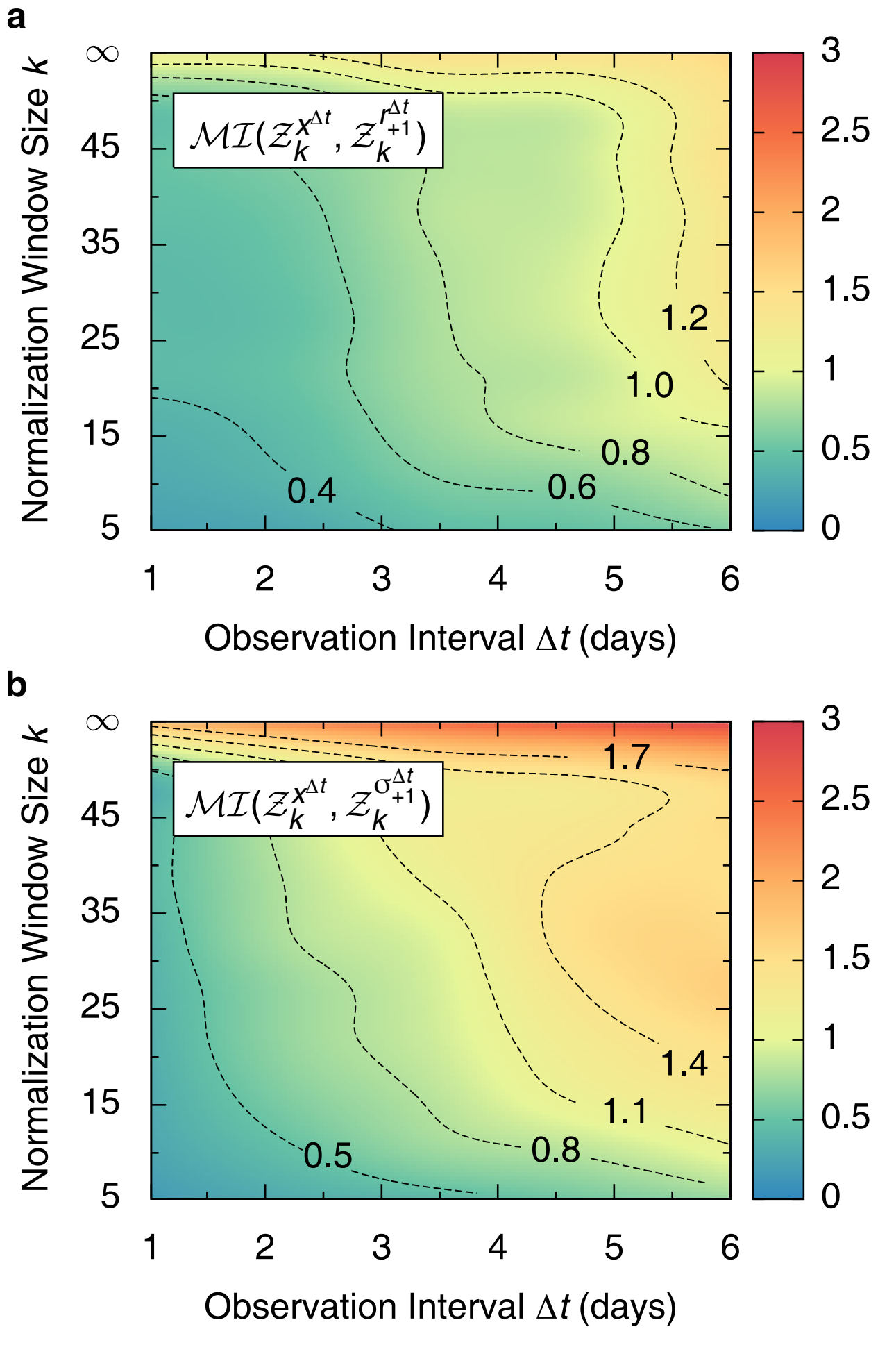

Figure 2: Mutual Information for Different Observation and Normalization Schemes. **a** mutual information between the input vector and returns; **b** mutual information between the input vector and volatility. The color scheme and contour lines are interpolations of discrete data on an integer grid. The calculations are performed based on equ. 9, assuming conditional independence between individual input dimensions.

Finally, we determine the optimal scheme through

$$\left(\hat{\Delta t}, \hat{k}\right) = \text{argmax} \mathcal{MI}\left(\mathcal{Z}_k^{x^{\Delta t}}, \mathcal{Z}_k^{\sigma_{+1}^{\Delta t}}\right). \tag{10}$$

The maximum predictive power can be achieve with long observation interval. However, we are also limited by the number of total sampling points. To allow sufficient (over 1,000) data samples, we choose

$$\Delta t = 3 \text{ days}, \quad k = \infty. \tag{11}$$

It is worth noting that different metrics and scheme spaces can be used to replace equ. 10 for different specific problems. However, the methods of scheme selection used in this work should be widely applicable.

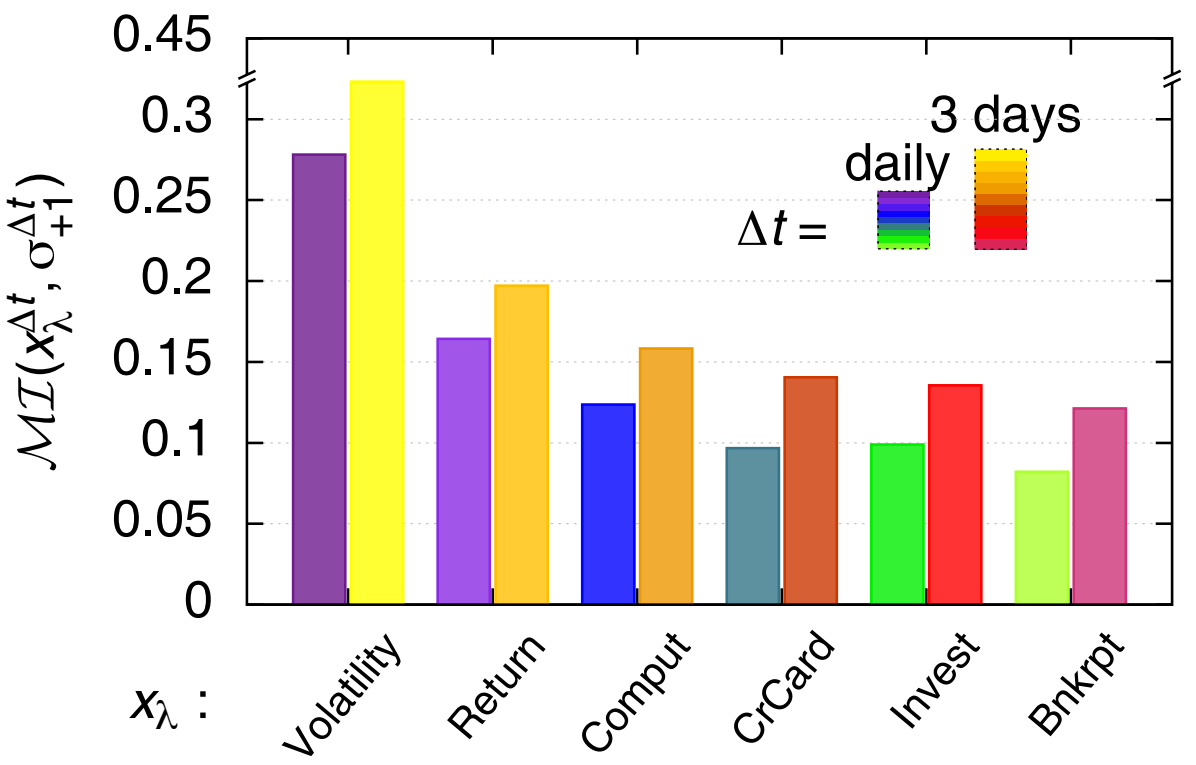

Figure 3: Individual Mutual Information for Different Input Components. Normalization window size $k = \infty$, observation interval $\Delta t = 1$ day (cold color) and 3 days (warm color) are used. Only the top 6 components with the strongest mutual information among all 25 input dimensions are shown in this plot.

Fig. 3 lists the highest mutual information for individual input components with $k = \infty$ and $\Delta t = 1$ day and $\Delta t = 3$ days. The lag-1 auto mutual information of volatility has the strongest predictive power for both observation time intervals. This is consistent with the observation that autoregressive time series models can be helpful at the daily timescale. Following the volatility itself, returns have the second strongest mutual information. The remaining components, which are Google domestic trends including computers & electronics, credit cards, finance & investing and bankruptcy, all have similar levels of predictive power. In later sections of the paper, we will drop the notation of $\Delta t$ and $k$ since their values are fixed for the rest of the analysis.

# Methods

In our recurrent neural network modeling of volatility, a single LSTM hidden layer consisting of one LSTM block is employed without other hidden layers. The structure of this neural network is shown in fig. 4. It has a dynamic "gating" mechanism. Running through the center is the cell state $I_i$ which we interpret as the information flow of the market sensitivity. $I_i$ has a memory of past time information [15] and more importantly it learns to forget [16] through equ. 12.

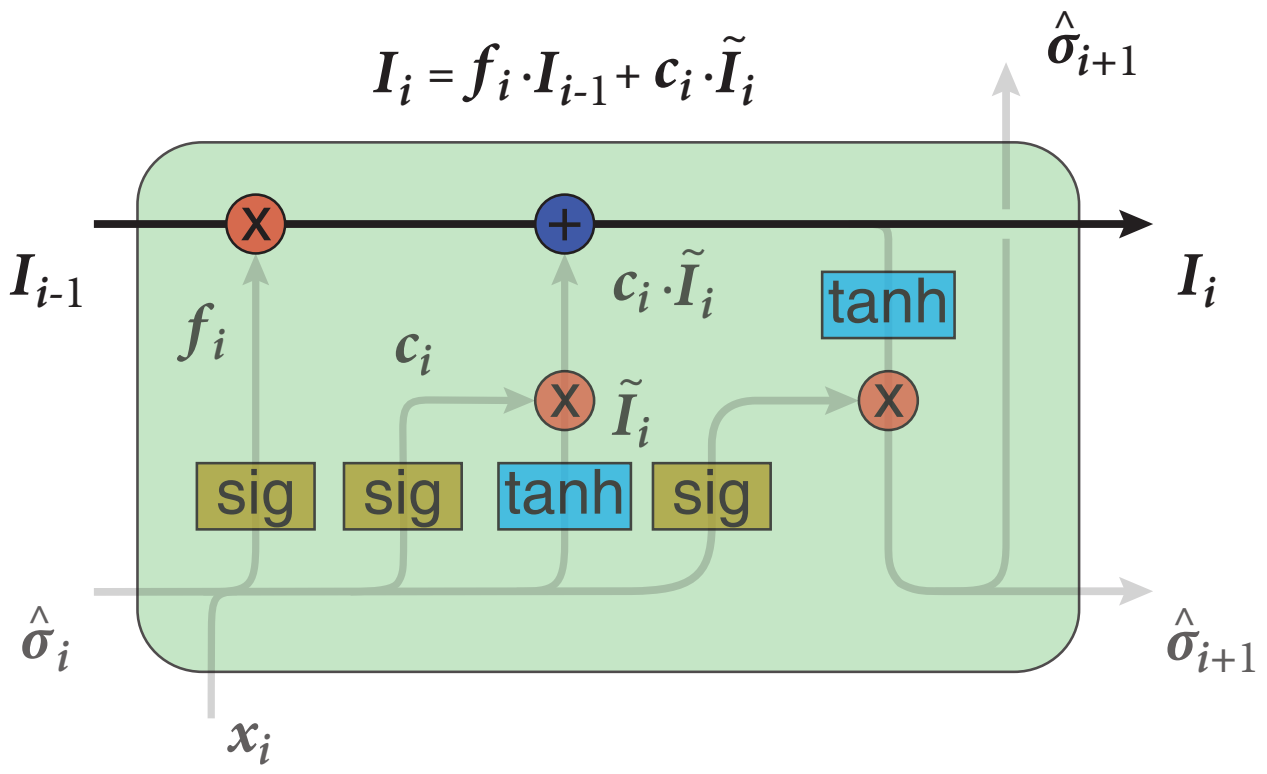

Figure 4: Structure of the Long Short-Term Memory Layer. At each time step $i$, this layer takes in input vector $x_{\lambda,i}$, volatility estimation $\hat{\sigma}_i$ and the information flow $I_{i-1}$ passed down from the last step. Gates are controlled by either the sigmoid (sig) function or the hyperbolic tangent (tanh) function while scalar multiplication and addition are denoted by $\times$ and $+$ operators. The linear memory update equation (equ. 12) is highlighted together with the cell state flow. Volatility prediction $\hat{\sigma}_{i+1}$ for the next time stamp and $I_i$ are passed down to the next epoch.

$$I_i = f_i \cdot I_{i-1} + c_i \cdot \tilde{I}_i \tag{12}$$

Here $f_i$ is the fraction of past-time information passed over to the present, $\tilde{I}_i$ measures the information flowing in at the current time and $c_i$ is the weight of how important this current information is. All these three quantities are functions of the input $x_{\lambda,i}$ and last-time's estimation of volatility $\hat{\sigma}_i$.

$$f_i = \text{sigmoid}\left[\left(\hat{\sigma}_i, x_i\right) \cdot W_f + b_f\right] \tag{13}$$

$$c_i = \text{sigmoid}\left[\left(\hat{\sigma}_i, x_i\right) \cdot W_c + b_c\right] \tag{14}$$

$$\tilde{I}_i = \tanh\left[\left(\hat{\sigma}_i, x_i\right) \cdot W_{\tilde{I}} + b_{\tilde{I}}\right] \tag{15}$$

To make a prediction of the next volatility value $\hat{\sigma}_{i+1}$, a linear activation function is used.

$$\hat{\sigma}_{i+1} = \alpha + \beta \cdot o_i \cdot \tanh\left[I_i\right] \tag{16}$$

Here $o_i$, which is also a function of $x_{\lambda,i}$ and $\hat{\sigma}_i$ tunes the output.

$$o_i = \text{sigmoid}\left[\left(\hat{\sigma}_i, x_i\right) \cdot W_o + b_o\right] \tag{17}$$

$I_i$ and $\hat{\sigma}_{i+1}$ are passed down to the next time step for continual predictions. Equ. 12 answers the fundamental question of memory in time series forecasting. Auto-regressive moving average model (ARMA($p$,$q$)) [17], however, answers this question by evaluating autocorrelation and partial autocorrelation functions and setting up the $p$ and $q$ maximum lags.

All coefficients here are learned through training with the python deep learning library Keras [18]. Specifically, we set up the maximum lag of the LSTM to include 10 successive observations, consistent with the benchmark linear models which we will describe below. The model is trained by the "Adam" method [19] with 32 examples in a batch and with mean absolute percent error (MAPE) as the objective loss function. 20% of the training data is held out to create a validation set, used in per-epoch error reporting. We have found that tuning the batch size and the validation fraction will change the MAPE in the test set by < 2% once the MAPE in the training set has reached 20% during the training process. This can be achieved after roughly 600 epochs. Moreover, data points are shuffled during training, no dropout has been implemented in our work [20] and all initial weights are set to be small positive constant terms, similar to the normalized initialization given in ref. [21].

To evaluate the performance of the LSTM model, 30% of the observed data is used as the test set. Additionally, we have developed two linear regression models (Ridge and Lasso) and one autoregressive model (GARCH) [17] as benchmark models.

$$\text{GARCH:}\ \sigma_i^2 = \omega + \sigma_{i-1}^2\left[\alpha + \beta\varepsilon_i^2\right],\ \varepsilon \sim \mathcal{N}(0,1) \tag{18}$$

$$\text{Linear:}\ \sigma_i = \omega + \varepsilon_i + \sum_{\lambda}\sum_{j=1}^{10}\alpha_{\lambda,j}x_{\lambda,i-j},\ \varepsilon \sim \mathcal{N}(0,\cdot) \tag{19}$$

While the GARCH model is easily trained by a maximum likelihood estimator, the linear models are regularized by $L_p$ norm of the coefficients $\alpha_{\lambda,j}$ thus giving two linear regression benchmarks: Lasso ($p = 1$) and Ridge ($p = 2$). More specifically, we set up a grid of regularization parameter $C$ from $10^{-2}$ to $10^{-6}$ spaced equally in the log scale and then train all of them on the first 80% of the training set by minimizing the following objective function.

$$O_p = C \cdot \left|\alpha_{\lambda,j}\right|_p + \sum \varepsilon_i^2 \tag{20}$$

The linear coefficients are determined using the later 20% validation part of the training set. We observe that the coefficients in volatility, return, bnkrpt, invest, and jobs are significantly non-zero in the linear Ridge model. This is consistent with the predictive power of each component as evaluated by the mutual information metric.

## Results

In fig. 5, we plot the predicted volatility together with the observed values in the test set. The subplot shows two types of error metric for our LSTM model, compared with the benchmark models. The MAPE is used as the loss function in training the neural network. Therefore, the LSTM has significantly lower MAPE (> 31% relatively) than any other benchmark models. In terms of root mean square error (RMSE), the LSTM also outperforms other benchmark models. However, the improvement is not as significant as on the MAPE.

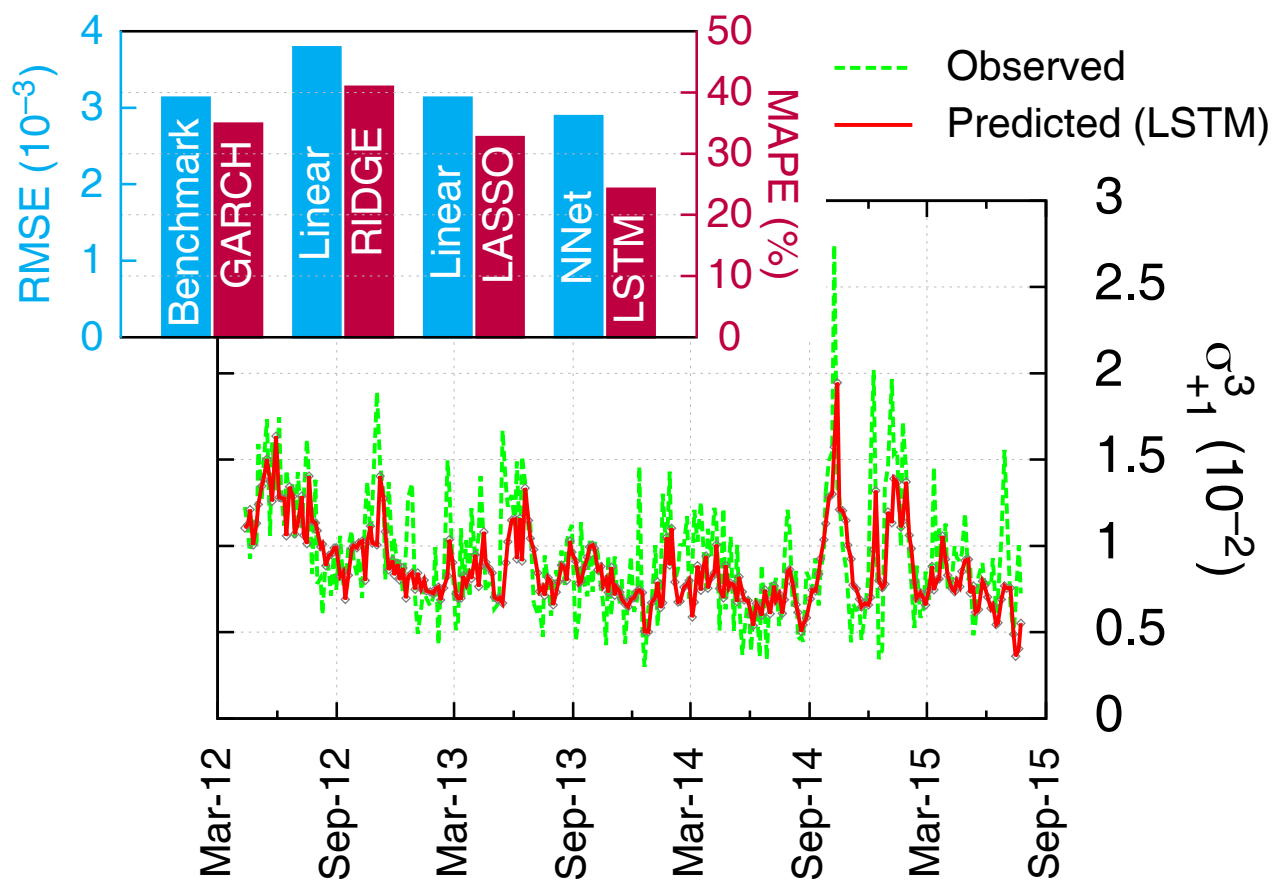

Figure 5: Volatility Forecasting Made by the Long Short-Term Memory Model and Comparison with Benchmarks. Volatility predictions (red) in the test set starting from 12-Apr-2012 to 24-Jul-2015 are presented with volatility observed at a $\Delta t = 3$ days timescale within the same time range (green). More specifically, this $\sigma^3$ (both prediction and observation) is the per-daily quadratic variation of the returns estimated with equ. 2 aggregated on a 3 day timescale with equ. 6. The subplot shows the comparison of the LSTM model with other benchmark models in two different error metrics, root mean square error (cyan) and mean absolute percentage error (purple).

Our LSTM model seems to avoid significant overfitting in the sense that the MAPE in the training set converges to roughly the same value (20%) as the MAPE evaluated in the test set (24.2%). We have further investigated overfitting by reducing the dimensionality of the input vector. Let us denote the LSTM with the full input $x_\lambda$ as $\text{LSTM}_0$. Let $\text{LSTM}_r$ be the one which has only a subset of the input vector as listed in fig. 3 including volatility, return, comput, crcard, invest and bnkrpt.

| Model | RMSE | MAPE* |
|---|---|---|
| $\text{LSTM}_0$ | $2.89 \times 10^{-3}$ | 24.2% |
| $\text{LSTM}_r$ | $2.88 \times 10^{-3}$ | 27.2% |
| Garch | $3.13 \times 10^{-3}$ | 34.9% |

Table 2: Error Metrics Evaluated in the Test Set. $\text{LSTM}_0$ is the original LSTM model and $\text{LSTM}_r$ is the reduced input dimension LSTM model. The Garch model, i.e. equ. 18, is also listed for comparison.

Table 2 shows that the test set MAPE in the reduced input model $\text{LSTM}_r$ increases from the original model $\text{LSTM}_0$.

## Discussion

The low signal-to-noise level poses a great challenge to all attempts trying to model the stock market at an intermediate timescale. Going to either a longer or shorter observation interval, one would have stronger deterministic patterns (see fig. 2) or auto-correlation. Still we choose this timescale to start our investigation so that the results presented in this work could be reproduced with publicly available data.

At a higher frequency, macroeconomic factors and public interest represented by the Google domestic trends will become less helpful in the prediction task. Presumably, one could use the market micro-structure instead as the input of the neural network, *e.g.* bid/ask prices and volume of the first few levels in the order book [22]. The input feature set, observation interval, and normalization scheme can be determined by maximizing some objective metrics (*e.g.*, mutual information) that best fit the specific problem. Moreover, news analytics [23] and arrival dynamics [24] could be of particular interest given their non-linear nature and neural networks' ease of expressing non-linearities.

For studies at similar frequencies, variations of our model can be applied with different inputs, stocks in different industries or different financial products, and different structures of the LSTM layer. There are two important improvements that can be made on this preliminary study: visualization of the hidden information flow and a confidence interval of the prediction. In this work, we understand the information flow $I_i$ as some hidden market state. It is not directly observable from the market quantities and is thus hard to be cross-validated. Further investigation of a direct connection between $I_i$ and other market observables could yield greater impact in financial time series modeling. In addition, statistics about the prediction error could help determine the confidence interval of the forecast. The most direct approach to obtain this knowledge is to evaluate the distribution and autocorrelation of the prediction error in the test set. In this work, we only observe that this error has zero mean ($< 10^{-5}$) and a standard deviation of $2.89 \times 10^{-3}$ (see fig. 5 and table 2). We have investigated the autocorrelation and partial autocorrelation functions of the error series in the test set. None of the lags are significant with respect to a $\pm 2$ standard deviation band. This shows that the prediction error has no memory of itself, as expected. However, this residual fails the one-sample Kolmogorov-Smirnov test [25] with a p-value $\ll 1\%$ for the null hypothesis that the residual is normally distributed. We feel that a deeper understanding of the prediction error is an essential task to be undertaken.

## Conclusion

In this work, we consider the Google domestic trends as environmental variables. Together with the market information, they constitute the driving force of daily S&P 500 volatility change. By constructing an appropriate mutual information metric, we find the optimal observation and normalization scheme for volatility forecasting. Within this scheme, we develop a neural network model which consists of one single long short-term memory layer and is trained on 70% of the entire data set. This model gives a MAPE of 24.2% in the remaining 30% of testing data, outperforming other linear and autoregressive benchmark models by at least 31%. This work shows the potential of deep learning financial time series in the presence of strong noise. The methods demonstrated in this work can be directly applicable for other financial quantities at completely different timescales where either correlation or deterministic drift outweigh noises.

## References

[1] K. Hornik, M. Stinchcombe and H. White, Multilayer Feedforward Networks Are Universal Approximators, Neural Networks **2**, 359 (1989).

[2] K. Kamijo and T. Tanigawa, Stock price pattern recognition—a recurrent neural network approach, IJCNN **1**, 215 (1990).

[3] S. A. Hamid and Z. Iqbal, Using Neural Networks for Forecasting Volatility of S&P 500 Index Futures Prices, J. Bus. Res. **10**, 1116 (2004).

[4] J. Bollen, H. Mao, and X. Zeng, Twitter Mood Predicts the Stock Market, J. Comput. Sci. **1**, 1 (2011).

[5] Z. Huang, H. Chen, C. Hsu, W. Chen and S. Wu, Credit Rating Analysis with Support Vector Machines and Neural Networks, A Market Comparative Study, Decis. Support Syst. **37**, 543 (2004).

[6] V. Nair, G. Hinton, Rectified Linear Units Improve Restricted Boltzmann Machines, Proceedings of the 27th International Conference on Machine Learning (2010).

[7] A. Mordvintsev, C. Olah, and M. Tyka, Inceptionism: Going Deeper into Neural Networks, Google Research Blog (2015).

[8] A. Graves, Generating Sequences With Recurrent Neural Networks, arXiv:1308.0850 (2013).

[9] M. Sundermeyer, R. Schluter and H. Ney, LSTM Neural Networks for Language Modeling, International Conference on Spoken Language Processing, InterSpeech (2010). 2010.

[10] A. Graves, A. Mohamed, and G. Hinton. Speech Recognition with Deep Recurrent Neural Networks, Acoustics, Speech and Signal Processing (ICASSP) (2013).

[11] M. B. Garman and M. J. Klass, On the Estimation of Security Price Volatilities from Historical Data, J. Bus. **1**, 67 (1980).

[12] T. Preis, H. S. Moat and H. E. Stanley, Quantifying Trading Behavior in Financial Markets Using Google Trends, Sci. Rep. **3**, 1684 (2013).

[13] S. E. Said and D. A. Dickey, Testing for Unit Roots in Autoregressive-Moving Average Models of Unknown Order, Biometrika **71**, 599–607 (1984).

[14] C. D. Manning, P. Raghavan and H. Schütze, An Introduction to Information Retrieval, Cambridge University Press (2008).

[15] S. Hochreiter and J. Schmidhuber, Long Short-Term Memory. Neural Comput **9**, 1735–1780 (1997).

[16] F. Gers, J. Shmidhuber and F. Cummmins, Learning to Forget: Continual Prediction with LSTM. Neural Comput **12**, 2451-2471 (2000).

[17] T. L. Lai and H. Xing, Statistical Models and Methods for Financial Markets, Springer (2008).

[18] F. Chollet, Keras, https://github/fchollet/keras, GitHub Repository (2015).

[19] D. Kingma and J. B. Adam, A Method for Stochastic Optimization, 3rd International Conference on Learning Representations (2015).

[20] N. Srivastava, G. Hinton, et al. Dropout: A Simple Way to Prevent Neural Networks from Overfitting. Journal of Machine Learning Research (JMLR) **15**, 1929-1958 (2014).

[21] X. Glorot and Y. Bengio, Understanding the Difficulty of Training Deep Feedforward Neural Networks, 13th International Conference on Artificial Intelligence and Statistics (2010).

[22] F. Abergel, J. P. Bouchaud, T. Foucault, C. A. Lehalle and M. Rosenbaum, Edited: Market Microstructure: Confronting Many Viewpoints, Wiley (2012).

[23] G. Mitra and L. Mitra, The Handbook of News Analytics in Finance, Wiley (2011).

[24] J. M. Maheu and T. H. Mccurdy, News Arrival, Jump Dynamics, and Volatility Components for Individual Stock Returns, J. Financ. **2**, 755 (2004).

[25] H. W. Lilliefors, On the Kolmogorov-Smirnov Test for Normality with Mean and Variance Unknown, J. Amer. Statist. Assoc. **62** 318 (1967).